\documentclass[journal=jpcafh,manuscript=article,layout=twocolumn]{achemso}

\usepackage{graphicx}
\usepackage[table-align-text-post=false, range-units=single, range-phrase=\textendash, separate-uncertainty=false, multi-part-units=single, per-mode=symbol-or-fraction]{siunitx}
\usepackage{threeparttable}
\usepackage{booktabs}
\usepackage[section]{placeins}

\usepackage{hyperref}
\usepackage{lineno}

\usepackage[version=4]{mhchem}
\newcommand{\refsuppinfo}{Supporting Information}

\author{Luis Bonah}
\author{Stephan Schlemmer}
\affiliation[Univsersity Cologne]
{I. Physikalisches Institut, Universität zu Köln, Zülpicher Str. 77, 50937 Köln, Germany}
\author{Jean-Claude Guillemin}
\affiliation[University Rennes]{Univ Rennes, Ecole Nationale Supérieure de Chimie de Rennes, CNRS, ISCR – UMR6226, 35000 Rennes, France}
\author{Michael E. Harding}
\affiliation[KIT]{ Institut f\"ur Nanotechnologie, Karlsruher Institut f\"ur Technologie (KIT), Kaiserstr. 12, 76131 Karlsruhe, Germany}
\author{Sven Thorwirth}
\email{sthorwirth@ph1.uni-koeln.de}
\affiliation[Univsersity Cologne]
{I. Physikalisches Institut, Universität zu Köln, Zülpicher Str. 77, 50937 Köln, Germany}

\title[Millimeter- and Submillimeter-Wave Study of C$_2$H$_5$CP]
  {On the Spectroscopy of Phosphaalkynes:\\ Millimeter- and Submillimeter-Wave Study of C$_2$H$_5$CP}


\begin{document}

\begin{tocentry}

\includegraphics[width=\textwidth]{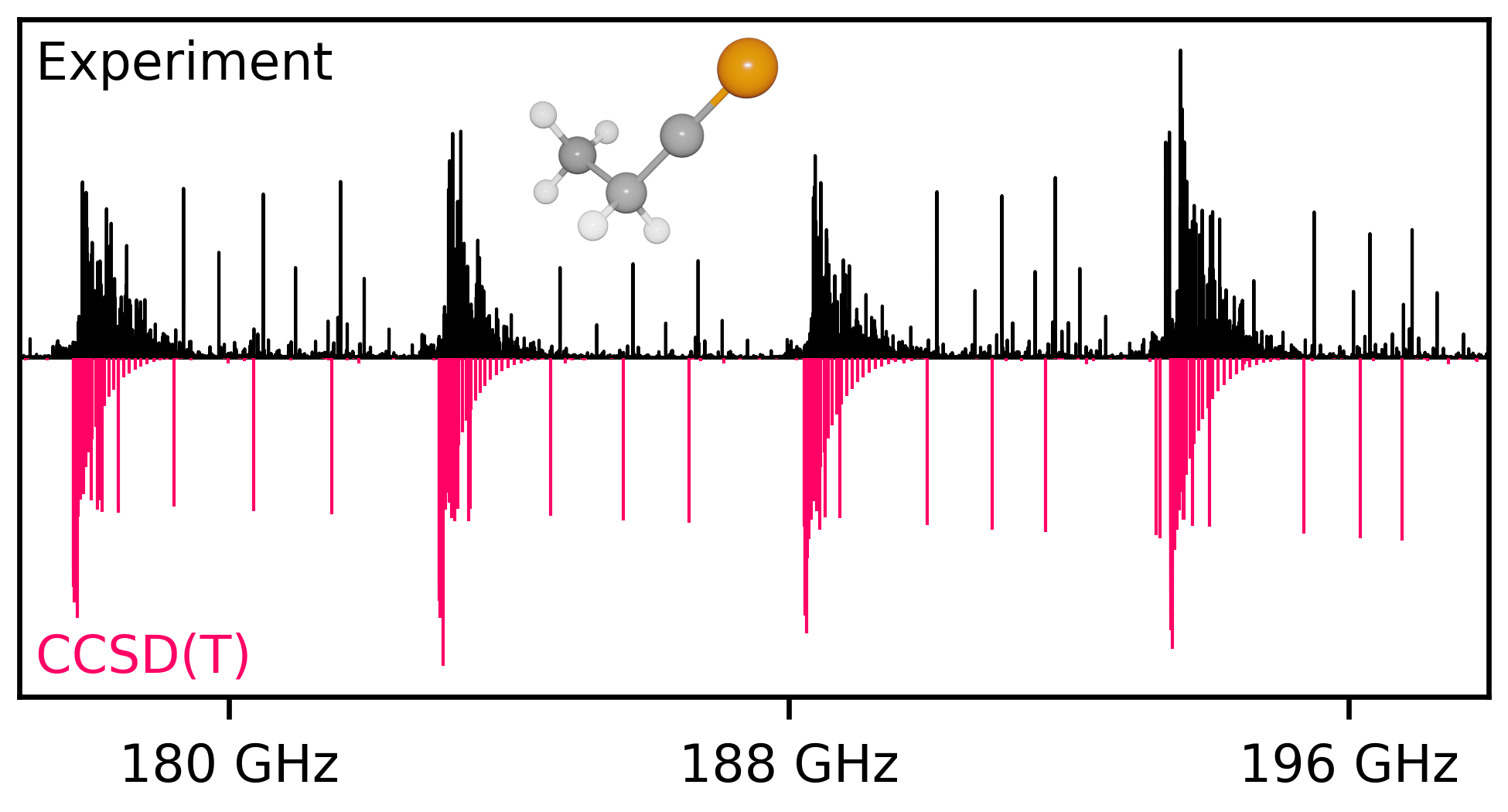}

\end{tocentry}

\begin{abstract}
  Ethyl phosphaethyne, \ce{C2H5CP}, has been characterized spectroscopically in the gas phase for the
  first time, employing millimeter- and submillimeter-wave spectroscopy 
  in the frequency regime from 75 to 760 GHz. Spectroscopic detection and analysis was guided by
  high-level quantum-chemical calculations of molecular structures and force fields performed at the coupled-cluster singles and doubles level extended by a perturbative correction for the contribution from triple excitations, CCSD(T), in combination with large basis sets. Besides the parent isotopologue, the three singly substituted \ce{^13C} species were observed in natural abundance up to frequencies as high as 500\,GHz. 
  Despite the comparably low astronomical abundance of phosphorus, phosphaalkynes, R--CP, such as 
  \ce{C2H5CP} are promising candidates for future radio astronomical detection. 
\end{abstract}

\section{Introduction}
Nitriles, chemical species of the general formula \ce{R-CN}, are one if not the most prominent class of molecules found in space.
Over the last fifty years, many nitriles have been detected by their pure rotational spectra using radio astronomical techniques.
Here, species range from simple prototypical \ce{HCN}~\cite{snyder_ApJL_163_L47_1971}, over metal-bearing variants like
\ce{FeCN}~\cite{zack_ApJ_733_L36_2011} and CN-bearing molecular ions~\cite{cernicharo_AAA_641_L9_2020,cernicharo_AAA_672_L13_2023}
up to complex and heavy benzenoid variants such as
cyanonaphthalene~\cite{mcguire_science_371_1265_2021} and cyanoindene~\cite{sita_ApJL_938_L12_2022}, just to name a few. 

Owing to intrinsically strong dipole moments, numerous nitriles have also been studied in the laboratory using microwave and millimeter-wave spectroscopy.
In contrast,
comparable studies of phosphaalkynes, where the \ce{CN} functional group is replaced by an isovalent \ce{CP} unit, are rather scarce, which may in part be attributed to their pronounced transient character and also more challenging synthetic routes. Since the first microwave spectroscopic
studies of prototypical phosphaethyne (\ce{HCP})~\cite{tyler_JCP_40_1170_1964}, and a handful of other selected species studied by Kroto and collaborators (\ce{HC3P}, \ce{CH3CP}, \ce{NCCP}, \ce{C2H3CP}, \ce{PhCP}; see \citet{{BurckettStLaurent_JMSt_79_215_1982}} and references therein), only a small number of additional phosphaalkynes have been characterized
employing high-resolution (rotational) spectroscopy. 
For detailed accounts on the available laboratory spectroscopic data of individual species, the interested reader may consult the reports on 
phosphaethyne, \ce{HCP}~\cite{bizzocchi_JMS_110_205_2001,bizzocchi_CPL_408_13_2005},
\ce{HC3P}~\cite{bizzocchi_CPL_319_411_2000,bizzocchi_JMS_110_205_2001},
\ce{HC5P}~\cite{bizzocchi_JCP_119_170_2003,bizzocchi_PCCP_5_4090_2003},
\ce{NCCP}~\cite{bizzocchi_JMS_110_205_2001},
\ce{NC4P}~\cite{bizzocchi_JMS_221_186_2003,bizzocchi_PCCP_6_46_2004},
\ce{C2H3CP}~\cite{ohno_JMS_90_507_1981},
\ce{CH3CP}~\cite{Transue_JACS_17985_140_2018,degliesposti_JMSt_1203_127429_2020},
\ce{PhCP}~\cite{burckettstlaurent_JMS_92_158_1982}, and
\ce{C3H5CH2CP}~\cite{samdal_JPCA_118_9994_2014}.

It should be noted that while to this day spectroscopic signatures of only
two phosphaalkynes, \ce{HCP} and \ce{NCCP}, have been found in space~\cite{agundez_ApJ_662_L91_2007,agundez_AAA_570_A45_2014} not all of the above species have yet been characterized in the laboratory at a level meeting the needs of radio astronomy.
In addition, other potentially astronomically relevant phosphaalkynes still await laboratory (high-resolution) spectroscopic characterization. One such species is ethyl phosphaethyne, \ce{C2H5CP} (also known as propylidynephosphine or phosphabutyne).

\ce{C2H5CP}, the phosphorus variant of the astronomically ubiquitous ethyl cyanide, \ce{C2H5CN} (e.g.~\citet{endres_JMS_375_111392_2021}),
has been known in the laboratory for many years
and characterized by its \ce{^1H}, \ce{^31P}, and \ce{^13C} NMR spectra in solution~\cite{guillemin_Angewandte_30_196_1991,guillemin_JOC_66_7864_2001}.
However, so far, there does not seem to be any account of its spectroscopic properties in the gas phase.
In the present study, high-level quantum-chemical calculations were performed at the coupled-cluster (CC) singles and doubles level extended by a perturbative correction for the contribution from triple excitations, CCSD(T).
Based on these results, the pure rotational spectrum of \ce{C2H5CP} has been detected for the first time and observed in selected frequency
ranges between \SIrange[range-phrase=\text{ and },range-units=single]{75}{760}{GHz}. A detailed account of the experimental and theoretical work as well as the analysis of the rotational spectrum of \ce{C2H5CP} in its ground vibrational state will be given in the following.

\section{Theoretical and Experimental Methods}

\subsection{Quantum-Chemical Calculations}
\label{sec:QuantumChemicalCalculations}

Quantum-chemical calculations to guide the spectroscopic search of \ce{C2H5CP} were performed at the CC singles and doubles level extended by a perturbative correction for the contribution from triple excitations (CCSD(T))~\cite{raghavachari_chemphyslett_157_479_1989}. All calculations were performed using the quantum-chemical program package CFOUR \cite{cfour,cfour_JCP_2020,harding_JChemTheoryComput_4_64_2008}. Correlation-consistent polarized valence and polarized core valence basis sets were used throughout. 
Within the frozen core (fc) approximation, the tight-$d$-augmented basis set cc-pV(T+$d$)Z was used for the phosphorus atom, and the corresponding cc-pVTZ basis sets for carbon and hydrogen \cite{dunning_JCP_90_1007_1989,dunning_JCP_114_9244_2001} as well as the atomic natural orbital basis set ANO1 \cite{almlof_JCP_86_4070_1987}. The ANO1 set consists of 18s13p6d4f2g to 5s4p2d1f, 13s8p6d4f2g to 4s3p2d1f, and 8s6p4d3f to 4s2p1d contractions for P, C, and H, respectively.

   \begin{figure}
   \centering
   \includegraphics[width=0.8\linewidth]{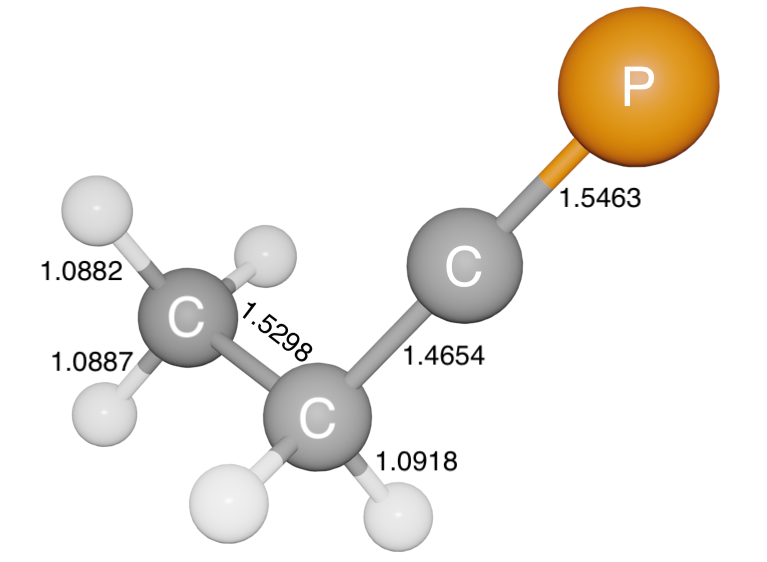}
      \caption{Bond lengths of \ce{C2H5CP} calculated at the ae-CCSD(T)/cc-pwCVQZ level of theory (in \AA ). Full structure in internal coordinates is given in the \refsuppinfo.
              }
         \label{FigVibStab}
   \end{figure}

The cc-pwCVXZ (X = T and Q) basis sets were used when considering all electrons in the correlation treatment \cite{peterson_JCP_117_10548_2002}. Equilibrium geometries were obtained using analytic gradient techniques \cite{watts_chemphyslett_200_1-2_1_1992}. For molecules comprising first and second-row elements, the ae-CCSD(T)/cc-pwCVQZ level of theory has been shown on many occasions to yield molecular equilibrium structures of very high quality (e.g., \citet{coriani_JCP_123_184107_2005}). The corresponding structure of \ce{C2H5CP} is shown in \autoref{FigVibStab}. Full sets of internal coordinates of both the ground state molecular structure and the transition state to methyl internal rotation are collected in the \refsuppinfo.  

Harmonic and anharmonic force fields were calculated in the fc approximation using the cc-pV(T+$d$)Z and ANO1 basis sets using analytic second-derivative techniques \cite{gauss_chemphyslett_276_70_1997,stanton_IntRevPhysChem_19_61_2000}, followed by additional numerical differentiation to calculate the third and fourth derivatives needed for the anharmonic force field \cite{stanton_IntRevPhysChem_19_61_2000,stanton_JCP_108_7190_1998}.
Theoretical ground-state rotational constants were then estimated from the equilibrium rotational constants (calculated at the CCSD(T)/cc-pwCVQZ level of theory) and the zero-point vibrational corrections $\Delta A_0$, $\Delta B_0$, and $\Delta C_0$ (calculated at the fc-CCSD(T)/ANO1 level, see \autoref{tab:MolecularParameters}). 
In addition, the fc-CCSD(T)/ANO1 force field calculation yields the quartic and sextic centrifugal distortion parameters. Spin-rotation constants, owing to the presence of \ce{^31P} ($I=1/2$), were calculated at the CCSD(T)/cc-pwCVQZ level of theory using rotational London orbitals \cite{gauss_JCP_105_2804_1996}. 

At the ae-CCSD(T)/cc-pwCVQZ level of theory and corrected for harmonic zero-point effects at the fc-CCSD(T)/ANO1 level of theory, 
the barriers of internal rotation $V_3$ of both \ce{C2H5CP} and isovalent
\ce{C2H5CN} are estimated as about 2.8\,kcal/mol, suggesting that torsional
splitting will not be significant for the ground state rotational spectrum obtained here.

\citet{puzzarini_IntRevPhysChem_29_273_2010} may be consulted for further insight into the theoretical approaches used in the present context.

\begin{table}[tb]
    \small
	\centering
	\caption{Calculated and experimental molecular parameters of \ce{C2H5CP} in its ground vibrational state.}
	\label{tab:MolecularParameters}

\resizebox{\linewidth}{!}{
	\begin{threeparttable}
\begin{tabular}{l l S[table-format=-5.3, round-mode = uncertainty] S[table-format=-5.8(2), round-mode = uncertainty]}
\toprule
  \multicolumn{2}{l}{Parameter}  &  \text{Calculations} &  \text{Experimental} \\
\midrule
     $A_e$ & / $ \si{\mega\hertz}  $ &    25374.531         &        $\cdots$       \\ 
     $B_e$ & / $ \si{\mega\hertz}  $ &      2719.663        &        $\cdots$        \\
     $C_e$ & / $ \si{\mega\hertz}  $ &      2532.995        &        $\cdots$        \\
$\Delta A_0$&/ $ \si{\mega\hertz}  $ &       158.770        &        $\cdots$        \\
$\Delta B_0$&/ $ \si{\mega\hertz}  $ &        12.603        &        $\cdots$        \\
$\Delta C_0$&/ $ \si{\mega\hertz}  $ &        14.132        &        $\cdots$        \\
     $A_0$ & / $ \si{\mega\hertz}  $ &     25215.761        &     25216.12285(35)   \\
     $B_0$ & / $ \si{\mega\hertz}  $ &      2707.060        &     2709.143447(22)   \\
     $C_0$ & / $ \si{\mega\hertz}  $ &      2518.863        &     2520.638536(21)   \\
  $-D_{J}$ & / $ \si{\hertz}       $ &      -879.142        &       -900.7695(38)   \\
 $-D_{JK}$ & / $ \si{\kilo\hertz}  $ &        23.760        &       24.101222(85)   \\
  $-D_{K}$ & / $ \si{\kilo\hertz}  $ &      -536.855        &       -547.7691(34)   \\
   $d_{1}$ & / $ \si{\hertz}       $ &      -147.802        &       -153.7099(23)   \\
   $d_{2}$ & / $ \si{\hertz}       $ &        -7.414        &        -8.11625(97)   \\
   $H_{J}$ & / $ \si{\milli\hertz} $ &         1.461        &         1.50407(31)   \\
  $H_{JK}$ & / $ \si{\milli\hertz} $ &       -14.470        &        -14.8371(54)   \\
  $H_{KJ}$ & / $ \si{\hertz}       $ &        -1.374        &        -1.37747(40)   \\
   $H_{K}$ & / $ \si{\hertz}       $ &        36.623        &          36.895(11)   \\
   $h_{1}$ & / $ \si{\micro\hertz} $ &       469.5          &          490.70(24)   \\
   $h_{2}$ & / $ \si{\micro\hertz} $ &        65.5          &           69.37(13)   \\
   $h_{3}$ & / $ \si{\micro\hertz} $ &         6.8          &           6.570(15)   \\
   $L_{J}$ & / $ \si{\nano\hertz}  $ &       $\cdots$       &         -3.0130(87)   \\
  $L_{JK}$ & / $ \si{\micro\hertz} $ &       $\cdots$       &          -2.446(11)   \\
 $L_{KKJ}$ & / $ \si{\micro\hertz} $ &       $\cdots$       &           71.50(53)   \\
   $l_{1}$ & / $ \si{\nano\hertz}  $ &       $\cdots$       &         -1.3244(74)   \\
   $l_{2}$ & / $ \si{\pico\hertz}  $ &       $\cdots$       &         -259.3(5.0)   \\
   $C_{aa}$(P)& / $ \si{\kilo\hertz}  $ &      7.42         &        $\cdots$        \\
   $C_{bb}$(P)& / $ \si{\kilo\hertz}  $ &      5.02         &        $\cdots$        \\
   $C_{cc}$(P)& / $ \si{\kilo\hertz}  $ &      4.92         &        $\cdots$        \\
   $C_{ab}$(P)& / $ \si{\kilo\hertz}  $ &     12.51         &        $\cdots$        \\
   $C_{ba}$(P)& / $ \si{\kilo\hertz}  $ &      1.06         &        $\cdots$        \\
   $\mu_a$ & / $ \si{D}            $ &      1.53            &        $\cdots$        \\
   $\mu_b$ & / $ \si{D}            $ &      0.29            &        $\cdots$        \\   
   $V_3$   & / $ \si{kcal/mol}     $ &      2.8             &                        \\
\midrule
\multicolumn{2}{l}{Transitions}   &    $\cdots$          &           6016  \\
\multicolumn{2}{l}{Lines}         &    $\cdots$          &           4010  \\
\textit{RMS} &/ \si{\kilo\hertz}  &    $\cdots$          & 29.7 \\ 
\multicolumn{2}{l}{\textit{WRMS}} &    $\cdots$          &   1.00   \\ 
\bottomrule
\end{tabular}
\begin{tablenotes}
\item \textbf{{Notes.}} Fits performed with \texttt{SPFIT} in the $S$-reduction and the $\text{I}^\text{r}$ representation.
Standard errors are given in parentheses. Lines that were rejected from the fit are excluded from the statistics.
Equilibrium rotational constants, phosphorus spin rotation constants, dipole moment components, and barrier of internal rotation calculated at the ae-CCSD(T)/cc-pwCVQZ level, zero-point vibrational contributions to the rotational constants, centrifugal distortion constants, and barrier of internal rotation calculated at the fc-CCSD(T)/ANO1 level. Ground state rotational constants are estimated as $B_0 = B_e - \Delta B_0$. For further details, see text.
\end{tablenotes}
\end{threeparttable}}
\end{table}

\subsection{Experiment}
Broadband measurements were recorded with a synthesized sample using three different experimental setups in Cologne.
The three measured frequency ranges of \SIrange{75}{120}{GHz}, \SIrange{170}{255}{GHz}, and \SIrange{340}{760}{GHz} resulted in a total frequency coverage of \SI{550}{GHz}.

\subsubsection{Synthesis}
\label{subsec:Synthesis}
\ce{C2H5CP} was prepared following the synthesis published by \citet{guillemin_JOC_66_7864_2001}. The three-step sequence involves the synthesis of 1,1-dichloropropylphosphonic acid diisopropyl ester which was reduced to 1,1-dichloropropylphosphine, followed by a bis-dehydrochlorination using 1,8- diazabicyclo[5,4,0]undec-7-ene as a base to generate \ce{C2H5CP}.
The two last steps of the synthesis were performed in tetraethylene glycol dimethyl ether (tetraglyme) as solvent. \ce{C2H5CP} was purified by vaporization, condensed under vacuum at \SI{-100}{\celsius}, and finally introduced into a cell containing degassed tetraglyme for storage at dry ice temperature.

\subsubsection{Broadband Measurements}
\label{subsec:BroadbandMeasurements}

Spectra were obtained through broadband scans by utilizing 
three different absorption experiments sharing a common basic structure consisting of a source, an absorption cell, and a detector.
The sources consist of synthesizers and subsequent commercial amplifier-multiplier chains to reach the desired frequency range.
For the absorption cells, different lengths of borosilicate tubes were used.
The resulting radiation was propagated via horn antennas, mirrors, and polarization filters through the absorption cells and onto the detectors.
These are either Schottky detectors ($<\SI{500}{GHz}$) or a cryogenically cooled bolometer ($>\SI{500}{GHz}$).
All experiments utilized frequency modulation with a \textit{2f}-demodulation scheme to increase the SNR\@.
The resulting lineshapes look similar to the second derivative of a Voigt profile.
More in-depth descriptions of the setups have been given earlier in \citet{martin-drumel_JMS_307_33_2015,Zingsheim2021}.

All measurements were performed at room temperature and static pressure.
This was deemed safe as a time series of the same calibration line showed no significant degradation of the sample over time.
Before each filling of the absorption cells, the sample was frozen out.
Then, \ce{C2H5CP} was selectively removed from the solution \textit{in vacuo} while keeping the sample container at a temperature of about \SI{-40}{\celsius}.
The filling pressures were in the range of \SIrange{10}{40}{\micro\bar}.
Reproducibility lines were measured before and after each batch of measurements as sanity checks (see \refsuppinfo). 
Standing waves were removed from the broadband spectra via Fourier filtering with an in-house written script\footnote{Available at \texttt{\url{https://github.com/Ltotheois/SnippetsForSpectroscopy/tree/main/FFTCorrection}}}.

\section{Results and Discussion}

\begin{figure*}[bth]
    \centering
    \includegraphics[width=\linewidth]{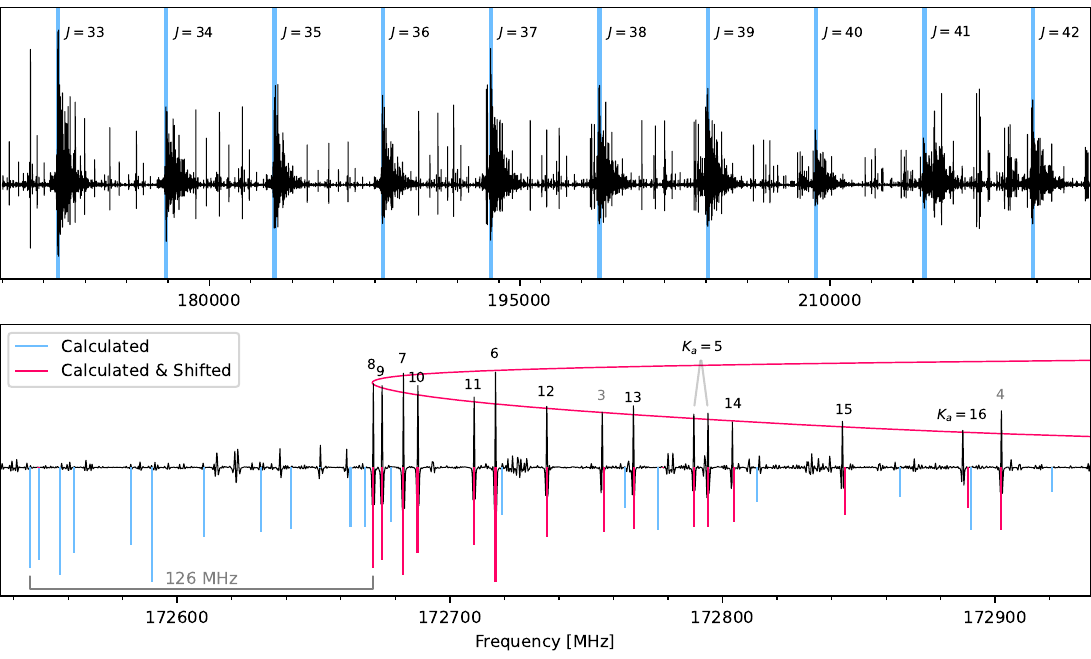}
    \caption{
        Top: Pure rotational spectrum of \ce{C2H5CP} from 170 to 216\,GHz.
        The typical pattern determined by consecutive $a-$type 
        rotational transitions is clearly visible.
        Transitions of the $J_{7,J-7} \leftarrow J-1_{7,J-8}$ series are highlighted in blue and the respective $J$ values are given.
        The intensities vary due to the frequency-dependence of the source power and pressure in the cell increasing over time.        
        Bottom: One of the first scans taken at \SI{172.7}{GHz}.
        The initial theoretical predictions (blue sticks) reproduce the experimental pattern nearly quantitatively when shifted by about +\SI{+126}{MHz},
        making spectroscopic assignment a straightforward procedure.
        The red trendline highlights the pattern of the $33_{K_a, 33-K_a} \leftarrow  32_{K_a, 32-K_a}$ and $33_{K_a, 34-K_a} \leftarrow  32_{K_a, 33-K_a}$ transitions.
        For $K_a > 5$ the asymmetry components are overlapping whereas the splitting is still resolved for $K_a=5$ (indicated by the gray lines).
        For $K_a = 3$ and $K_a=4$ the second asymmetry components lie outside the shown region.
    }
    \label{fig:InitialMeasurement}
\end{figure*}

\ce{C2H5CP} is an asymmetric top molecule with Ray's asymmetry parameter of $\kappa = (2B-A-C)/(A-C) = \SI{-0.98}{}$. %
Thus, \ce{C2H5CP} is a highly prolate rotor close to the symmetric limit of \SI{-1}{}.
Its two nonzero dipole moments $\mu_a = \SI{1.5}{D}$ and $\mu_b = \SI{0.3}{D}$ (see \autoref{tab:MolecularParameters}) result in a strong $a$-type spectrum and a considerably weaker $b$-type spectrum with the
methyl group potentially allowing for internal rotation splitting and energetically low-lying vibrational states giving rise to rather intense vibrational satellite spectra.

\subsection{Parent Isotopologue}
\label{sec:ParentIsotopologue}

Initial scans of the millimeter-wave regime were performed around a wavelength of \SI{2}{mm}
in search of the characteristic $a$-type line pattern.
Indeed, strong transitions were observed soon and were found to be fully compatible
with the theoretical predictions on both large and small frequency scales. As can be seen in the top spectrum of \autoref{fig:InitialMeasurement}, 
the broadband scan around \SI{200}{GHz} clearly features harmonically related line series, as expected for the $a$-type spectrum of a prolate
asymmetric rotor. From coarse visual inspection of this scan alone, the separation between consecutive transitions is about \SI{5}{GHz}, in very good agreement
with the theoretical estimate of $(B+C)=\SI{5.25}{GHz}$ calculated for \ce{C2H5CP} (\autoref{tab:MolecularParameters}). At smaller scales, see the bottom spectrum in \autoref{fig:InitialMeasurement},
the assignment of \ce{C2H5CP} to the carrier of the molecular absorption is secured immediately:
As can be seen, %
the theoretical spectrum can be brought into near-perfect agreement with the dominant transitions of the experimental spectrum when a small shift of only \SI{+126}{MHz} is applied, identifying the experimental lines as belonging to the $J=33 \leftarrow 32$ transition.
The spectroscopic offset of \SI{126}{MHz} translates into a deviation of merely $\SI{126}{MHz}/33\approx \SI{4}{MHz}$ in $B+C$ between prediction and experiment, corresponding to an effective agreement on the order of one per mill.
This small empirical correction permitted immediate detection and assignment of many lines
from adjacent rotational transitions over a sizable quantum number regime. %

Comprehensive spectroscopic analysis was finally
carried out using the \SIrange{75}{120}{GHz}, \SIrange{170}{255}{GHz},
and \SIrange{340}{760}{GHz} broadband spectra %
and by tracing line series in Loomis-Wood plots employing
the LLWP software \cite{bonah_JMS_388_111674_2022}.
Besides the capabilities of LLWP to facilitate spectroscopic assignment and (multicomponent) spectral
line profile fitting to evaluate transition frequencies accurately, 
it has also been designed as a frontend to Pickett's \texttt{SPFIT}/\texttt{SPCAT} program suite \cite{pickett_JMolSpectrosc_148_371_1991}, hence speeding up the spectroscopic analysis as a whole significantly.
Owing to the high quality and predictive power of the CCSD(T) model (\autoref{tab:MolecularParameters}),
spectral assignment was a rather straightforward procedure.
Furthermore, consistency of line fitting was evaluated and supported using
our Python package \textit{Pyckett} which is a Python wrapper around Pickett's \texttt{SPFIT} and \texttt{SPCAT} program suite adding
some very useful functionality\footnote{
This work made extensive use of Pycketts CLI tools \texttt{pyckett\_add} and \texttt{pyckett\_omit} to analyze the influence of adding sensible additional parameters to the Hamiltonian or omitting any of the included parameters from the Hamiltonian.
See \texttt{\url{https://pypi.org/project/pyckett/}} for more information or install with pip via \texttt{pip install pyckett}}.
Rather than estimating uniform transition frequency uncertainty,
uncertainties of either \SI{20}{kHz}, \SI{30}{kHz}, or \SI{40}{kHz} were assigned via an automated process which is described in greater detail in the \refsuppinfo.

The final fit comprises \SI{6041}{} ground state transitions with \SI{4024}{} unique frequencies
spanning quantum number ranges of $4 \le J \le 140$ and $0\le K_a \le 25$. 
The majority of these lines are $a$-type transitions (\SI{4340}{} transitions with \SI{2856}{} unique frequencies) complemented with $b$-type transitions (\SI{1701}{} transitions with \SI{1223}{} unique frequencies) that proved more difficult to assign due to their lower intensities. 
However, LLWP's Blended Lines Window allowed us to derive accurate line positions even for weak or moderately blended lines.
No $A/E$-splitting from internal rotation of the methyl group was observed.
The final fit parameters are 
collected in \autoref{tab:MolecularParameters}. As can be seen, the full set of sextic and five octic centrifugal
distortion parameters were needed to reproduce the transition frequencies within their experimental uncertainties ($rms$ = \SI{30}{kHz}, $wrms$ = \SI{1.00}{}).
The agreement between the calculated and the experimental molecular parameters is excellent.

\begin{figure}
    \centering
    \includegraphics[width=\linewidth]{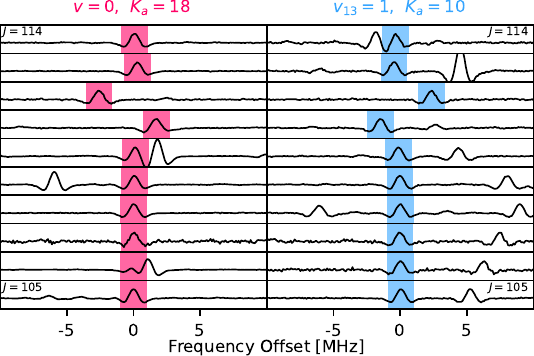}
    \caption{Side-by-side Loomis-Wood plots for the $J_{18, J-17} \leftarrow J-1_{18, J-18}$ series of the vibrational ground state $v=0$ (left side in red) and the $J_{10, J-9} \leftarrow J-1_{10, J-10}$ series of the energetically lowest vibrationally excited state $v_{13}=1$ (right side in blue).
    The two sides are almost perfect mirror images with strong deviations for $J=111$ and $J=112$.
    This indicates interactions between the two states centered around the respective $J=111$ energy levels.
    The predictions from this work were used as the center frequencies for $v=0$ while for $v_{13}=1$ a polynomial of degree 2 was fitted to the ten here seen assignments as its preliminary analysis is not sufficiently accurate.
    }
    \label{fig:Interaction_LWP}
\end{figure}

Only a very small number of
\SI{25}{} transitions at \SI{14}{} unique frequencies 
were omitted from the fit due to $(\nu_\text{obs} - \nu_\text{calc}) / \Delta\nu_\text{obs}$ values greater than \SI{5}{}.
Clearly, at values of about $109\le J \le 114$, the ground vibrational state shows signs of a local perturbation, see \autoref{fig:Interaction_LWP}.
From a rudimentary Boltzmann-analysis performed through intensity comparison of vibrational satellites and
ground state lines, the vibrational wavenumber is estimated as \SI{170}{cm^{-1}} hinting toward the energetically lowest state $v_{13}=1$, the first excited in-plane \ce{C-C-P} bending mode (see \refsuppinfo). While a quantitative perturbation treatment is
beyond the scope of the present paper, a preliminary fit of the vibrational satellite lines compared
against the ground state parameters yields rotation-vibration interaction constants in good agreement with
values obtained from the anharmonic force field calculations
\footnote{Calculations yield $\alpha_{\nu_{13}}^{A} = \SI{146.32}{MHz}$, $\alpha_{\nu_{13}}^{B} = \SI{-8.28}{MHz} $, $\alpha_{\nu_{13}}^{C} =  \SI{-3.61}{MHz}$ while a preliminary fit yields $\alpha_{\nu_{13}}^{A} =  \SI{180(3)}{MHz}$, $\alpha_{\nu_{13}}^{B} =  \SI{-8.280(7)}{MHz}$, $\alpha_{\nu_{13}}^{C} =  \SI{-3.613(7)}{MHz}$.}, substantiating this assignment.
Further yet preliminary inspection of the $\nu_{13}$ vibrational satellite
pattern with LLWP provides evidence that the state is not just part of a dyad with the ground vibrational state but part of a polyad with other vibrational states. A comprehensive treatment of the vibrational satellite spectrum will be the subject of future analysis.

Additionally, $a$-type transitions of the vibrational ground state with $K_a = 22$ and $K_a \geq 26$ show systematic deviations for high $J$ values which are most-likely also a result of the interaction with the $\nu_{13}$ vibrational state. Thus the analysis of the parent isotopologue was limited to transitions with $K_{a} < 26$.

\subsection{Singly Substituted \ce{^{13}C} Isotopologues}
\label{sec:13CIsotopologues}

\begin{figure}
    \centering
    \includegraphics[width=\linewidth]{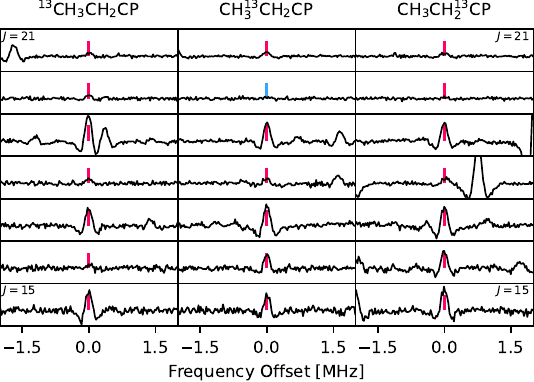}
    \caption{Loomis-Wood plot of the $J_{0, J} \leftarrow J-1_{0, J-1}$ transitions for the three singly substituted \ce{^{13}C} species of \ce{C2H5CP}.
    Predictions are shown as sticks in red if the transition is assigned or blue if unassigned.
    The measured intensity of the lines is highly dependent on the power of the source at the respective frequencies.
    The overall signal-to-noise ratio is very low for these isotopologues, severely limiting the number of possible assignments.
    }
    \label{fig:LWP_13C}
\end{figure}

While the detection of the singly substituted \ce{^{13}C} species 
proved challenging due to low line intensities as well as the overall high line density and line blending (see \autoref{fig:LWP_13C}), spectroscopic assignment was finally feasible
based on empirically improved predictions obtained by scaling the
theoretical rotational parameters of the minor species with correction factors obtained from 
the parent species (see \refsuppinfo). This procedure finally enabled the detection of $K_a=0,1$ line series very close to the scaled predictions using the LLWP program.
Once a valid spectroscopic assignment had been accomplished,
assignment of additional series was much easier, however, only $a-$type
spectra up to \SI{500}{GHz} were assigned with confidence.
Many transitions were found blended with close-by strong lines.
This was especially severe for \ce{CH3CH_{2\ \ }^{13}CP}, as its $B$ and $C$ constants are very similar to those of the parent isotopologue.
As a result, the quantum number coverage of \ce{CH3CH_{2\ \ }^{13}CP} is significantly more limited compared with that of the other two species.

\begin{table*}[h]
    \small
	\centering
	\caption{Molecular parameters of \ce{C2H5CP} and its singly substituted \ce{^{13}C} isotopologues.}
	\label{tab:MolecularParameters13C}

\resizebox{\textwidth}{!}{
\begin{threeparttable}
\centering
\begin{tabular}{l l *{ 4 }{S[table-format=-6.10]}}
\toprule
\multicolumn{2}{l}{Parameter} & \ce{CH3CH_2CP} & \ce{^{13}CH3CH_2CP} & \ce{CH_{3 \ \ }^{13}CH_2CP} & \ce{CH3CH_{2\ \ }^{13}CP} \\
\midrule
$ A                    $&/ $ \si{\mega\hertz}     $&    25216.12285(35)  &        24834.81(19)&        24696.03(15)&        25115.37(19)\\ 
$ B                    $&/ $ \si{\mega\hertz}     $&    2709.143447(22)  &      2641.07228(69)&      2686.01405(83)&      2708.97088(95)\\ 
$ C                    $&/ $ \si{\mega\hertz}     $&    2520.638536(21)  &      2457.88562(70)&      2495.32816(71)&      2519.45345(67)\\ 
$ -D_{J}               $&/ $ \si{\hertz}          $&      -900.7695(38)  &         -883.46(13)&         -861.16(15)&         -897.91(23)\\ 
$ -D_{JK}              $&/ $ \si{\kilo\hertz}     $&      24.101222(85)  &         24.6990(22)&         22.2244(20)&          24.467(13)\\ 
$ -D_{K}               $&/ $ \si{\kilo\hertz}     $&      -547.7691(34)  &            -552(19)&            -505(20)&            -563(18)\\ 
$ d_{1}                $&/ $ \si{\hertz}          $&      -153.7099(23)  &         -149.84(20)&         -149.67(24)&         -155.40(23)\\ 
$ d_{2}                $&/ $ \si{\hertz}          $&       -8.11625(97)  &          -7.635(45)&          -8.242(74)&          -8.133(43)\\ 
$ H_{J}                $&/ $ \si{\milli\hertz}    $&        1.50407(31)  &           1.507(25)&           1.363(25)&           1.582(45)\\ 
$ H_{JK}               $&/ $ \si{\milli\hertz}    $&       -14.8371(54)  &          -17.12(51)&          -11.89(45)&          -14.1(3.3)\\ 
$ H_{KJ}               $&/ $ \si{\hertz}          $&       -1.37747(40)  &         -1.3748(89)&         -1.3072(70)&           -1.34(11)\\ 
$ H_{K}                $&/ $ \si{\hertz}          $&         36.895(11)  &\text{a}&\text{a}&\text{a}\\ 
$ h_{1}                $&/ $ \si{\micro\hertz}    $&         490.70(24)  &             518(44)&             503(53)&             549(49)\\ 
$ h_{2}                $&/ $ \si{\micro\hertz}    $&          69.37(13)  &\text{a}&\text{a}&\text{a}\\ 
$ h_{3}                $&/ $ \si{\micro\hertz}    $&          6.570(15)  &\text{a}&\text{a}&\text{a}\\ 
$ L_{J}                $&/ $ \si{\nano\hertz}     $&        -3.0130(87)  &\text{a}&\text{a}&\text{a}\\ 
$ L_{JK}               $&/ $ \si{\micro\hertz}    $&         -2.446(11)  &\text{a}&\text{a}&\text{a}\\ 
$ L_{KKJ}              $&/ $ \si{\micro\hertz}    $&          71.50(53)  &\text{a}&\text{a}&\text{a}\\ 
$ l_{1}                $&/ $ \si{\nano\hertz}     $&        -1.3244(74)  &\text{a}&\text{a}&\text{a}\\ 
$ l_{2}                $&/ $ \si{\pico\hertz}     $&        -259.3(5.0)  &\text{a}&\text{a}&\text{a}\\ 
\midrule
\multicolumn{2}{l}{Transitions}         &   6016    &   417   &  556    &  163    \\
\multicolumn{2}{l}{Lines}               &   4010    &   282   &  366    &  133    \\
\textit{RMS} &/ \si{\kilo\hertz}         & 29.7     & 26.5    & 25.3    & 27.7    \\ 
\multicolumn{2}{l}{\textit{WRMS}}       & 1.00      & 0.75    & 0.71    & 0.77    \\ 
\bottomrule
\end{tabular}
\begin{tablenotes}\footnotesize
\item \textbf{{Note.}} Fits performed with \texttt{SPFIT} in the S-reduction and the $\text{I}^\text{r}$ representation.
Standard errors are given in parentheses.
    \item \tnote{a} Parameter was fixed to the parent isotopologue value.
\end{tablenotes}
\end{threeparttable}
}
\end{table*}

For all rare isotopologues, full quadratic and quartic parameter sets were used.
For the sextic parameters only $H_J$, $H_{JK}$, $H_{KJ}$, and $h_1$ were determined which is a result of having only $a$-type transitions assigned.
To provide more accurate predictions outside the covered quantum number range, undetermined parameters were set to the main isotopologue values.
The resulting molecular parameters are listed in \autoref{tab:MolecularParameters13C}.
The parameters match the scaled predictions very well (see \refsuppinfo).

While the isotopic data obtained in the present study are by far not sufficient to derive an unconstrained empirical molecular structure of \ce{C2H5CP} they may be used
to derive structural information about the carbon backbone.
If the experimental ground state rotational constants of all four isotopologues available are first corrected for the effects of zero-point vibration (calculated here at the fc-CCSD(T)/ANO1 level of theory, \autoref{tab:MolecularParameters} and \refsuppinfo) prior to structural refinement,
then semi-experimental equilibrium structural parameters $r_{\rm e}^{\rm SE}$ are obtained \cite{vazquez_equilibrium_structures_2011,Kisiel2003} that may be compared directly to their \textit{ab initio} values. Using this strategy and keeping the majority
of structural parameters fixed at their ae-CCSD(T)/cc-pwCVQZ values, the two C--C bond lengths 
as well as the C-C-C-angle are determined as $r_{{\rm H}_3C-C{\rm H}_2}=1.5293(2)$\,\AA , $r_{{\rm H}_2C-C\rm P}=1.4647(2)$\,\AA
, and $\alpha_{CCC}=112.20(1)^{\circ}$. These values are in excellent agreement with their
ae-CCSD(T)/cc-pwCVQZ counterparts (see \refsuppinfo).
Any extension of the empirical structure determination in the future will require the experimental characterization of a much larger sample of isotopologues, most notably deuterated variants.

\section{Conclusions}
Ethyl phosphaalkyne, \ce{C2H5CP}, has been detected spectroscopically in the gas phase for the first time.
The pure rotational spectrum of the parent isotopic species could be detected, assigned, and analyzed covering frequencies as high as 760\,GHz. In addition, the three singly substituted \ce{^13C}-species were also observed and characterized up to 500\,GHz.
The experimental findings agree very well with the results of high-level CCSD(T) calculations.

Future analysis of the vibrational satellite spectrum will be an interesting and challenging task. \ce{C2H5CP} possesses a sizable number of energetically low-lying vibrational modes. Consequently, the vibrational satellite spectrum will comprise contributions not only from fundamental vibrations but also from overtone and combination modes
additionally providing ample opportunity for interactions and hence perturbed spectra. As has been shown here, even the ground vibrational state is subject to perturbation at high values of $J$ and $K_a$.

Now that the pure rotational spectrum of \ce{C2H5CP} in the ground vibrational state has been studied to high accuracy, astronomical searches for this phosphaalkyne in suitable sources are feasible.
Frequency predictions along with all relevant data (line lists, fit files) will be provided and archived through the Cologne Database for Molecular Spectroscopy, CDMS \cite{mueller_cdms,endres_cdms_2016}.

\begin{suppinfo}
    Minimum and transition state structure as well as vibrational wavenumbers and rotation-vibration interaction constants resulting from the quantum-chemical calculations.
    The calculated and scaled rotational constants for the singly \ce{^{13}C} substituted isotopologues.
    More in detail descriptions of the reproducibility measurements and the automated transition frequency uncertainty assignment procedure.
\end{suppinfo}

\begin{acknowledgement}
LB, SS, and ST gratefully acknowledge the Collaborative Research Center 1601 (SFB 1601 sub-project A4) funded by the Deutsche Forschungsgemeinschaft (DFG, German Research Foundation) – 500700252.

      MEH acknowledges support by the Bundesministerium für Bildung und Forschung (BMBF) through the Helmholtz research program ``Materials Systems Engineering'' (MSE).

      JC-G thanks the CNRS national program PCMI (Physics and Chemistry of the Interstellar Medium) and the University of Rennes for a grant.
\end{acknowledgement}

\bibliography{sthorwirth_bibdesk}

\end{document}